\documentclass[12pt]{article}
\usepackage{hyperref}
\usepackage{graphicx}
\usepackage{amsmath}
\usepackage{amsthm}
\usepackage{amssymb}
\usepackage{braket}
\usepackage{tipa}
\usepackage{subfig} 
\usepackage{xcolor}
\usepackage{mathrsfs}
\usepackage{url}

\usepackage{empheq}
\theoremstyle{plain}

\usepackage{tikz}
\usepackage{tikz-cd}
\usetikzlibrary{arrows}
\usetikzlibrary{intersections}
\usetikzlibrary{shapes.geometric}
\usetikzlibrary{decorations.markings,bending}
\usetikzlibrary{decorations.pathreplacing,angles,quotes}

\usepackage[nosort]{cite}

\newlength{\abstractwidth}
\renewcommand{\title}[1]{\vbox{\center\bf{\Large{#1}}}\vspace{5mm}}
\renewcommand{\author}[1]{\vbox{\center#1}\vspace{5mm}}
\newcommand{\address}[1]{\vbox{\center\em#1}}
\newcommand{\email}[1]{\vbox{\center\tt#1}\vspace{5mm}}

\renewcommand{\bar}{\overline}
\renewcommand{\tilde}{\widetilde}

\renewcommand{\Re}{\operatorname{Re}}

\newcommand{\Tr}{\operatorname{Tr}}
\newcommand{\nn}{\nonumber}

\newcommand{\OTOC}{\text{OTOC}}
\newcommand{\NOC}{\text{NOC}}

\newcommand{\calC}{\mathcal{C}}

\newcommand{\calM}{\mathcal{M}}

\begin{document} 
	
\begin{titlepage}

\begin{center}
\hfill \\
\hfill \\
\vskip 1cm
	
\title{Out-of-Time-Order Correlation Functions for Unitary Minimal Models}
	
\author{Ruihua Fan}
\address{Department of Physics, Harvard University, Cambridge, MA 02138, USA}
\email{ruihua\_fan@g.harvard.edu}
	
\end{center}	
	
\begin{abstract}
	
	We analytically study the Out-of-Time-Order Correlation functions (OTOC) for two spatially separated primary operators in two-dimensional unitary minimal models. Besides giving general arguments using the conformal symmetry, we also use the Coulomb gas formalism to explicitly calculate the OTOC across the full time regime. In contrast to large-$N$ chaotic systems, these models do not display a separation of time scales, due to the lack of a large parameter. We find that the physics at early times ($0<t-x\ll\beta$) and late times ($t-x\gg\beta$) are controlled by different OPE channels, which are related to each other via the braiding matrix. The normalized OTOC obeys a power-law decay with a fractional power at early times and approaches a generically nonzero value in an exponential way at late times. The late time value is related to the modular $S$-matrix and is in agreement with earlier calculations\cite{rcft1,rcft2}. All of the results above are readily generalized to rational conformal field theories.
		
\end{abstract}

\end{titlepage}

\tableofcontents

\section{Introduction}
\label{sec:introduction}

Scrambling describes the process where local information under unitary time evolution gets lost in a many-body system and cannot be retrieved by local measurements. This concept, first discussed in black hole problems\cite{hayden,sekino}, is now also becoming more and more important for understanding other systems. Quantitatively, scrambling can be described by the out-of-time-order correlation function (OTOC)\cite{larkin, kitaev1},
\begin{align}
	C_1(t) = \frac{1}{Z}\Tr[\rho W^\dag(t)V^\dag(0) W(t)V(0)],
\end{align}
where $\rho=e^{-\beta H}$ is the thermal density matrix and $Z=\Tr\rho$ is the partition function. Here, $W$ and $V$ are two local operators and $W(t)=e^{itH}We^{-itH}$. An intuitive way to understand it is to consider the commutator square $\Tr[\rho [W(t),V(0)]^\dag [W(t), V(0)]]/Z$, whose growth tells us the expansion of local operators. When $W$ and $V$ are local unitary, it becomes $2 - 2\Re C_1(t)$. We can see that the decrease of the OTOC signals the expansion of local operators and thus the loss of local information. If the OTOC increases again after the decrease, it means the information travels in the system like a wavepacket, implying the underlying quasi-particles. If the OTOC keeps decreasing to a late-time value (smaller than $\braket{W^\dag W}_\beta\braket{V^\dag V}_\beta$), this is information scrambling, and the time scale for the OTOC decreasing significantly defines a scrambling time $t_{scr}$.

A related but different concept is thermalization\cite{srednicki,holographicmatter}, which describes how excitations under unitary time evolution collide and approach a thermal state. Although there are many criteria for various aspects of thermalization, locally it can be described by the normal-order correlation function (NOC),
\begin{align}
	C_2(t) = \frac{1}{Z} \Tr[\rho V^\dag(0)W^\dag(t) W(t)V(0)],
\end{align}
which can be understood as measuring $W^\dag(t) W(t)$ on a thermal state perturbed by a local operator $V$. For a generic interacting system, we expect thermalization to happen at large $t$ and measuring $W^\dag(t) W(t)$ gives the same result as doing measurement on a thermal state at the same temperature. Hence the NOC will approach the equilibrium value $\braket{W^\dag W}_\beta\braket{V^\dag V}_\beta$. This process defines another time scale, the dissipation time $t_{diss}$.

Both two correlation functions have been studied in many systems, including black holes\cite{bh1,bh2,bh3,channel}, SYK models\cite{SYK1,SYK2,SYK3,SYK4,sykliouville,SYK5,SYK6,SYK7,SYK8,SYK9,SYK10,SYK11,syk_kitaev,SYK12,SYK13}, interacting quantum field theories\cite{field1,prove,field2,rcft1,rcft2,field3,field4,field5,field6,field7,field8,field9}, many-body localized systems\cite{mbl1,mbl2,mbl3,mbl4}, random circuits\cite{circuit1,circuit2,circuit3}, random Hamiltonians\cite{randomH1,randomH2} and 1D lattice models \cite{lattice1,lattice2,lattice3,lattice4,lattice5}. It was realized that for some systems, the scrambling time is much larger than the dissipation time $t_{scr}\gg t_{diss}$. In the time regime $t_{diss}<t < t_{scr}$, the OTOC normalized by the NOC decays as $1-\frac{1}{N}e^{\lambda_L t}$, where $N\gg 1$ is some large parameter and $0<\lambda_L\le 2\pi/\beta$ is called the Lyapunov exponent\cite{prove}. These two features define a chaotic system in quantum many-body physics. Even for systems that do not satisfy these two criteria, studying these two quantities is still helpful in understanding the interplay between interactions and information spreading.

In spite of these extensive studies, the OTOC is mostly computed either in a specific time regime with approximations or with numerics. It would enhance our understanding to have analytical control of this quantity over the whole time regime. Two-dimensional conformal field theory provides such a platform, especially unitary minimal models, which will be the main focus of this paper. These models are simple to solve in principle but still contain strong enough interactions so that they don't have quasi-particles in general. 

In the rest of this paper, we will focus on unitary minimal models with infinite system size at finite temperature $1/\beta$. To partially get rid of the divergence of field theory, we are going to consider the following redefined OTOC and NOC,
\begin{align}
	\label{eqn:otoc}
	C_1(t) & = \frac{1}{Z}\Tr[\rho^{1/2}W(t)V(0)\rho^{1/2}W(t)V(0)], \\
	\label{eqn:no}
	C_2(t) & = \frac{1}{Z}\Tr[\rho^{1/2}V(0)W(t)\rho^{1/2}W(t)V(0)],
\end{align}
where $W$ and $V$ are chosen as Hermitian primary operators\footnote{
	In some literature, the OTOC is defined as $\Tr[\rho^{1/4}W(t)\rho^{1/4}V(0)\rho^{1/4}W(t)\rho^{1/4}V(0)]$. We will justify the advantage of our choice in the following discussion.
}. 

This paper is organized as follows. In Sec.~\ref{sec:preliminaries}, we will review some necessary facts and define notations about 2D CFTs and the Coulomb gas formalism of unitary minimal models. Readers familiar with these concepts can quickly go through them. We will then review the second-sheet effect, which leads to the difference between OTOCs and NOCs. In Sec.~\ref{sec:argument}, we will take advantage of the finiteness of primary fields to give a general argument about the behavior of the OTOC, NOC and their ratio $f(t)=C_1(t)/C_2(t)$ at different time regimes. In Sec.~\ref{sec:result}, we use the Coulomb gas formalism to explicitly compute the OTOC of certain operators for the whole time regime. Besides confirming our arguments, we also tune the central charge and see how the ratio changes. In Sec.~\ref{sec:conclusion}, we will give a summary and some additional remarks. Readers only interested in the main results can directly go to Sec.~\ref{sec:conclusion}. All the calculation details will be saved to Appendices~\ref{app:coulomb} and \ref{app:hypergeo}.

\section{Preliminaries}
\label{sec:preliminaries}

Although we need four point functions in real time, it is easier to handle CFTs in imaginary time. Thus our strategy is to first solve the four point function in imaginary time and then perform analytical continuation. In this section, we will discuss the general structure of Euclidean four point functions of 2D CFTs, in which models we have a full answer of the four point function and how to perform analytical continuation.

\subsection{Four point functions of 2D CFTs}

Both $C_1(t)$ and $C_2(t)$ can be generated by the following parent function which is defined on Euclidean plane,
\begin{align}
	\label{eqn:parent4pt}
	\braket{W(z_1,\bar z_1)W(z_2, \bar z_2)V(z_3, \bar z_3)V(z_4, \bar z_4)}.
\end{align}
Because $W$ and $V$ are primary operators, conformal invariance fixes it to the following form\cite{BPZ},
\begin{align}
	\braket{W(z_1,\bar z_1)W(z_2, \bar z_2)V(z_3, \bar z_3)V(z_4, \bar z_4)} = \frac{1}{z_{12}^{2h_W}z_{34}^{2h_V}\bar z_{12}^{2\bar h_W} \bar z_{34}^{2\bar h_V}} f_{WV}(z,\bar z).
\end{align}
where $z=z_{12}z_{34}/z_{13}z_{24}$ is the cross ratio. The prefactor is exactly $\braket{WW}\braket{VV}$. $f_{WV}(z,\bar z)$ contains dynamical data beyond conformal invariance and can be expanded in terms of the conformal blocks,
\begin{align}\label{eqn:fWV}
	f_{WV}(z,\bar z) = \sum_p F_p(z) \bar F_p(\bar z) = \sum_q \tilde F_q\left(\frac{1}{z}\right) \bar{\tilde F_q} \left( \frac{1}{\bar z} \right).
\end{align}
$p$ denotes the common fusion channels of $WW$ and $VV$ OPEs while $q$ denotes the fusion channels of $WV$ OPE. For a generic CFT, there may be infinite fusion channels, which makes calculation very difficult. For rational CFTs, the summations over $p$ or $q$ are always finite and the calculation becomes tractable. The two conformal blocks have the following series expansion,
\begin{align}
	F_p(z) &= z^{h_p} \sum_{\{K\}} A_p^{\{K\}} z^{K}, \\
	\tilde F_q\left(\frac{1}{z}\right) &= \left(\frac{1}{z} \right)^{h_q-h_W-h_V} \sum_{\{K\}} B_q^{\{K\}} \left(\frac{1}{z} \right)^K,
\end{align}
where $\{K\}$ represents the collection of all non-singular descendants in the corresponding conformal family and $K$ is the level of the descendants. Although the physical four point function is uniquely defined, conformal blocks can have branch cuts in general. Here, we choose the convention that $F_p(z)$ has a branch cut $[1,+\infty)$ while $\tilde F_q(1/z)$ has a branch cut $(-\infty, 1]$. We will see that $F_p(z)$ is useful for studying the late-time behavior and $F_q(1/z)$ is useful for the discussion of the early-time regime. The exact meaning of these time regimes will be properly defined later.

\subsection{Unitary minimal models and the Coulomb gas formalism}

Computing conformal blocks and four point functions in general CFTs is a formidable task. Here, we only consider unitary minimal models\cite{BPZ}. This is a series of models with only a finite number of primary fields. Each primary has infinite singular descendants. All these features make calculations tractable. Each model can be labeled by an integer $m\ge 3$ and denoted as $\calM(m+1,m)$. For $\calM(m+1,m)$, its central charge and the conformal weights of the primary fields $\phi_{r,s}$ take the following discrete values,
\begin{align}
	c & = 1-\frac{6}{m(m+1)}, \\
	h_{r,s} & = \frac{((m+1)r - ms)^2 - 1}{4m(m+1)},
\end{align}
where $1\le r\le m-1,1\le s\le m$ are two integers. As we can see, the central charge for these models is bounded above by $1$. $\calM(4,3)$ corresponds to the critical Ising model, while $m\rightarrow \infty$ gives free boson. Hence, although the central charge to some extent characterizes the number of degrees of freedom, among the minimal models, larger $c$ doesn't mean more chaotic. For simplicity, we will restrict ourselves to the diagonal series, so that every primary field is spinless.

There are different ways to construct the unitary minimal models. In this paper, we will use the Coulomb gas formalism\cite{coulombgas}, which can be understood as free bosons with a background charge $-2\alpha_0<0$ pinned at infinity. The central charge is reduced by the background charge to $c=1-24\alpha_0^2$. The primary fields are the vertex operators. Furthermore, we only consider those vertex operators whose charges and conformal weights satisfy the following formula,
\begin{align}
	\alpha_{r,s} = \frac{1}{2}(1-r)\alpha_+ + \frac{1}{2}(1-s)\alpha_-,\quad
	h_{r,s} = \frac{1}{4}(r\alpha_+ + s\alpha_-)^2 - \alpha_0^2
\end{align}
where $\alpha_\pm = \alpha_0 \pm \sqrt{\alpha_0^2 + 1}$ and $r,s$ are integers. These kind of vertex operators $V_{r,s}$ or $V_{-r,-s}$ can be identified with primary fields $\phi_{r,s}$ in the minimal models. Doing so, the correlators between $\phi_{r,s}$ can be translated into correlators for vertex operators which is much easier to compute. Readers interested in the details are referred to Appendix~\ref{app:coulomb}.

\subsection{Second-sheet effect}

Once we know the conformal blocks and the parent four point function Eqn\eqref{eqn:parent4pt}, we can translate it to finite temperature via the following mapping,
\begin{align}
	z = \exp\left(\frac{2\pi}{\beta}(x+i\tau)\right),\quad
	\bar z = \exp\left(\frac{2\pi}{\beta}(x-i\tau)  \right),
\end{align}
where $\tau$ is the imaginary time and $\beta=1/T$ is the inverse temperature. Here $z$ means the complex coordinate of each operator, not the cross ratio. To get real-time correlators, we have to analytically continue the imaginary time to a complexified time $t_c = t - i\tau$. The mapping is now written as,
\begin{align}
	z = \exp\left(\frac{2\pi}{\beta}(x - t_c)\right),\quad
	\bar z = \exp\left(\frac{2\pi}{\beta}(x + t_c)  \right).
\end{align}
Specifically, to get Eqn\eqref{eqn:otoc} and Eqn\eqref{eqn:no}, we have to assign,
\begin{align}\label{eqn:continuation}
	\rm{OTOC:}\, & \tau_1 \rightarrow \beta/2+\epsilon, \tau_2 \rightarrow \epsilon, \tau_3 \rightarrow \beta/2, \tau_4 \rightarrow 0, \\
	\rm{NOC:}\, & \tau_1 \rightarrow \beta/2, \tau_2 \rightarrow \epsilon, \tau_3 \rightarrow \beta/2+\epsilon, \tau_4 \rightarrow 0,
\end{align}
where $0<\epsilon\ll \beta$ is crucial to keep the relative ordering between $W$ and $V$. 

These two continuations do not look very different so one may simply expect the OTOC and NOC to behave similarly to each other. However, this is incorrect. The reason for that is the second-sheet effect\cite{field1}. 

\begin{figure}[!h]
	\centering
	\includegraphics[width=0.6\textwidth]{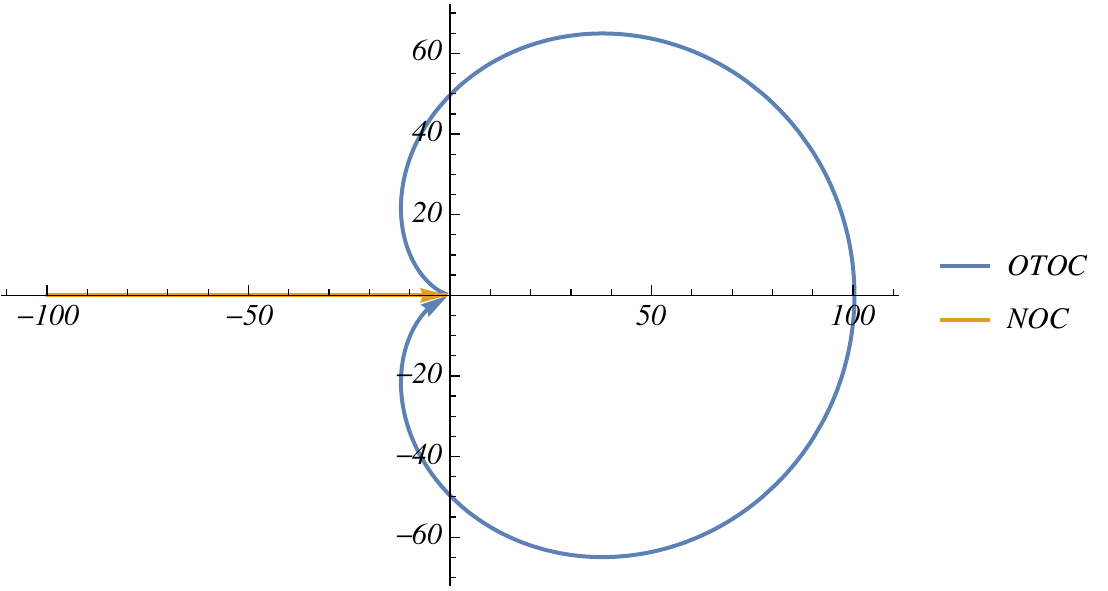}
	\caption{\label{fig:winding} Trajectories of $\bar z_\OTOC$ and $\bar z_\NOC$ when the real time $t$ increases from $0$ to a large enough value. $\bar z_\OTOC$ has a non-trivial winding around $\bar z=1$ while $\bar z_\NOC$ doesn't. We choose $\beta=2\pi$ and $\epsilon=0.1$ to make this effect clear to see on the plot. }
\end{figure}

Let's look at how cross ratio evolves under the analytical continuation. We put $W$ at $(0,t)$ and $V$ at $(x,0)$. Then given our continuation scheme Eqn\eqref{eqn:continuation}, the cross ratios are,
\begin{align}
	\textrm{OTOC :}& \,  z = \frac{-1}{\left[\sinh \frac{\pi}{\beta}(x+t-i\epsilon) \right]^2 },\, \bar z = \frac{-1}{\left[\sinh \frac{\pi}{\beta}(x-t+i\epsilon) \right]^2 }, \\
	\textrm{NOC :}& \, z =  \frac{2\cos^2 \frac{\pi\epsilon}{\beta}}{\cos\frac{2\pi\epsilon}{\beta} - \cosh \frac{2\pi(x+t)}{\beta}},\, \bar z = \frac{2\cos^2 \frac{\pi\epsilon}{\beta}}{\cos\frac{2\pi\epsilon}{\beta} - \cosh \frac{2\pi(x-t)}{\beta}}.
\end{align}
For $z$, we can safely take $\epsilon \rightarrow 0$ and find there is no difference between the OTOC and NOC and we don't make the distinction between them below. \footnote{
	Here we manually add a negative imaginary part to guarantee that $z$ sits below real axis so that $z$ is the complex conjugation of $\bar z$ at $t=0$.
}
\begin{align}
	z=z_\OTOC = z_\NOC = \frac{2}{1-\cosh \frac{2\pi(x+t)}{\beta}} - i\epsilon = \frac{-1}{\left[\sinh \frac{\pi}{\beta}(x+t) \right]^2 }- i\epsilon.
\end{align}
However, for $\bar z$, we see that the imaginary part of $\bar{z}_{\textrm{OTOC}}$ changes sign at $t=x$ while $\bar{z}_{\textrm{NOC}}$ doesn't. If we plot their trajectories on the complex plane, we can see that $\bar z_\OTOC$ winds around $\bar z=1$ while $\bar z_\NOC$ doesn't, as shown in Fig.~\ref{fig:winding}. 

Now it is easier to use $\bar F_p(\bar z)$ to analyze. Because the conformal blocks $\bar F_p(\bar z)$ have a branch cut at $[1,+\infty)$, such a winding will bring it to the second Riemann sheet. To be concrete, for the OTOC, when $\bar z$ goes across the branch cut to the second Riemann sheet, $\bar F_p(\bar z)$ has to pick up a monodromy matrix, i.e. $\bar F_p \rightarrow M_{pp'}\bar F_{p'}$, so that the OTOC and NOC have different values. If the $WW$ or $VV$ OPE only has a single fusion channel, then the monodromy matrix will be a phase factor. In this case, the OTOC and NOC only have a phase difference and there is no scrambling. In the following, we discuss generic cases so we assume there are multiple fusion channels and the monodromy matrix is not a phase factor.

One can also use $\bar{\tilde F}_q(1/\bar z)$, whose branch cut is at $(-\infty,1]$. Although $\bar z_\OTOC$ doesn't cross the branch cut of $\bar{ \tilde F}_q(1/\bar z)$, it turns out that the conformal blocks can still detect the phase difference between $\bar z_\OTOC$ and $\bar z_\NOC$, which leads to the difference between the OTOC and NOC, as we will discuss in detail below.

\section{General arguments}
\label{sec:argument}

In this section, we will combine the general structure of the four point functions and the second sheet effect to give some general arguments about OTOC at different time regimes. All of our arguments are mainly based on the fact that there is a finite number of primary fields in minimal models, so it is readily generalized to RCFTs\cite{rcft1,rcft2} but not general CFTs.

As we discussed before, the OTOC is different from the NOC because of the non-trivial winding of the right-hand cross ratio $\bar z$, which happens in a narrow time window (controlled by $\epsilon$) at the light cone $t=x$.\footnote{
	This actually depends on our definition of OTOC. If we choose another definition, for example, $\Tr[\rho^{1/4}W(t)\rho^{1/4}V(0)\rho^{1/4}W(t)\rho^{1/4}V(0)]$, $\bar z_\OTOC$ will start winding at the first beginning. So different choices will give different early time behaviors and our choice has the advantage of showing the causality clearly.	
} So given temperature and the spatial separation of $W$ and $V$, we can divide the time axis into different regimes: (1) $t<x$, $\bar z_\OTOC$ hasn't started winding so OTOC takes the same value as NOC. Physically this is because $V$ hasn't entered the light cone of $W(t)$. Neither scrambling nor thermalization happens thus we don't expect differences between OTOC and NOC. (2) $t>x$, OTOC starts to deviate from NOC. We will further divide it into early-time regime $\epsilon \ll t-x \ll \beta$ and late-time regime $t-x\gg \beta$, which we will discuss in detail below.\footnote{
	One can check that OTOC and NOC has the same prefactor $\braket{WW}_\beta\braket{VV}_\beta$, which is time-independent. So from now on we will drop the prefactor and only focus on $f_{WV}(z,\bar z)$. And what we mean by $C_1(t)$ and $C_2(t)$ is actually $f_{WV}(z,\bar z)$.
}

\subsection{Early time regime}

In this time regime, on the one hand we require $\epsilon \ll t-x$ so that $\bar{z}_{\textrm{OTOC}}$ has completed the winding and we can assume $\bar{z}_{\textrm{OTOC}}$ and $\bar{z}_{\textrm{NOC}}$ have the same real part but opposite imaginary part, or $\bar z_\OTOC = e^{-2\pi i} \bar z_\NOC$. On the other hand, because of $t-x \ll \beta$ we have $1/|\bar{z}| =-\frac{\pi^2}{\beta^2}(t-x)^2 \ll 1$. Therefore we can do a $1/\bar z$ expansion in the four point function, which inspires us to use the conformal blocks $\tilde F_q$. Thus the early-time physics is controlled by the $WV$ OPE. By dropping the unimportant prefactor, we have,
\begin{align*}
	C_1(t) = \sum_{q} \tilde F_q(1/z) \bar{\tilde F_{q}}(1/\bar z_{\OTOC}),\quad
	C_2(t) = \sum_{q} \tilde F_q(1/z) \bar{\tilde F_{q}}(1/\bar z_{\NOC}).
\end{align*}
The branch cut for $\bar{\tilde F}$ is $(-\infty,1]$ and $\bar z_\OTOC$ doesn't cross it so we don't add the monodromy matrix here. Suppose we sort and label the $WV$ fusion channels in a way that $h_{q_1} < h_{q_2} <...< h_{q_n}$. The leading terms of the $C_1(t)$ and $C_2(t)$ expansions are both proportional to $|\bar z|^{h_W+h_V-h_{q_1}}$, which indicates a light cone singularity because of $h_{q_1}-h_W-h_V<0$. To get rid of this singularity, we can instead compute their ratio $f(t)=C_1(t)/C_2(t)$. Noticing that $\bar z_\NOC/\bar z_\OTOC = e^{2\pi i}$, we have $F_{q}(1/\bar z_\OTOC) = F_{q}(1/\bar z_\NOC) e^{2\pi i(h_q-h_W-h_V)}$. So we can write the expansion of $C_1(t), C_2(t)$ as,
\begin{align*}
	C_1(t) \approx & \tilde F_{q_1}(1/z) \bar{\tilde F}_{q_1}(1/\bar z_\NOC) e^{2\pi i(h_{q_1}-h_W-h_V)} + F_{q_2}(1/z) \bar{\tilde F}_{q_2}(1/\bar z_\NOC) e^{2\pi i(h_{q_2}-h_W-h_V)}, \\
	C_2(t) \approx & \tilde F_{q_1}(1/z) \bar{\tilde F}_{q_1}(1/\bar z_\NOC) + F_{q_2}(1/z) \bar{\tilde F}_{q_2}(1/\bar z_\NOC),
\end{align*}
where we only preserve conformal blocks with the smallest two conformal weights. After a $1/\bar z$ expansion, the ratio is
\begin{align} \nn
	f(t) \approx & e^{2\pi i (h_{q_1}-h_W-h_V)} \left(1 + 2i\sin \left[ (h_{q_1}-h_{q_2})\pi \right] \frac{\tilde F_{q_2}(1/z) B_{q_2}}{\tilde F_{q_1}(1/z) B_{q_1}} |\bar z|^{h_{q_1}-h_{q_2}}+...
	\right) \\
	\approx & e^{2\pi i (h_{q_1}-h_W-h_V)} \left(1 + 2i\sin\left[ (h_{q_1}-h_{q_2})\pi \right] \frac{\tilde F_{q_2}(1/z) B_{q_2}}{\tilde F_{q_1}(1/z) B_{q_1}} \left(\frac{\pi^2}{\beta^2}(t-x)^2\right)^{h_{q_2}-h_{q_1}}+...
	\right)
\end{align}
The holomorphic cross ratio $z$ does not diverge in this time regime so $\tilde F_q(1/z)$ has a $O(1)$ value and its expansion with respect to $(t-x)$ only gives higher order contribution. As a result, we find the early-time behavior is $e^{2\pi i (h_{q_1}-h_W-h_V)}(1+\# (t-x)^r)$ with a fractional power $r=2(h_{q_2}-h_{q_1})$.

\subsection{Late time regime}

For the late-time regime $t-x\gg \beta$, we have $z,\bar z \sim e^{-\frac{2\pi}{\beta}t} \ll 1$. In this case, it will be easier to use the conformal blocks $F_p(z)$. So the late-time behavior is dominated by $WW$ and $VV$ OPE. With the consideration of second-sheet effect, We have,
\begin{align}
	C_1(t,x) = \sum_{p,p'} F_p(z) M_{pp'} \bar F_{p'}(\bar z),\,
	C_2(t,x) = \sum_{p} F_p(z) \bar F_{p}(\bar z).
\end{align}
Again we sort the fusion channels so that $h_{p_1} < h_{p_2} <...<h_{p_n}$, where $p_1$ is the identity channel. Therefore, we can directly set $h_{p_1}=0$. A similar analysis shows the ratio is,
\begin{align}
	f(t,x) \approx M_{p_1p_1} + \frac{A_{p_2}}{A_{p_1}}\left( (M_{p_1p_2}-M_{p_1p_1}) \bar z^{h_{p_2}} + (M_{p_2p_1}-M_{p_1p_1}) z^{h_{p_2}})  \right).
\end{align}
We find that the late-time behavior is $M_{p_1p_1}+\# \exp(-s\frac{2\pi}{\beta}t)$ with $s=h_{p_2}$ controlled by $WW$ and $VV$ OPE. The late-time value $M_{p_1p_2}$ has a relation with the modular $S$-matrix, as discussed in\cite{rcft1,rcft2}.

\paragraph{A few comments}
We see that $F_q(1/z)$ conveniently describes the early time while $F_p(z)$ is more appropriate for the late time. Because these two different conformal blocks are related by a braiding matrix, it means that if we know the behavior in one time regime, we can use the braiding matrix to predict the other one. 

We can see that, for both the early-time or the late-time expansions, the only time scale is $\beta$ , and there are no large parameters. Thus, we don't expect separation of time scales which we will verify below.

\section{Concrete examples}
\label{sec:result}

To be concrete, we explicitly compute and show the analytical result of OTOC and NOC in this section. For simplicity, we fix the choice of operators to be $W=\phi_{1,2}$ and $V=\phi_{m,n}$ in the following discussion.

\subsection{Parent four point functions}

In this section, we will list the results for the parent four point function without proof. Readers can refer to the appendix for details. We number the operators as $1$ to $4$ from left to right. Each operator is identified with a corresponding vertex operator with charge $\alpha_j$. Then the answer can be written as,
\begin{align}
	\braket{\phi_{(1,2)}\phi_{(1,2)}\phi_{(m,n)}\phi_{(m,n)}} = \frac{1}{z_{12}^{2h_w}z_{34}^{2h_v}} 
	\frac{1}{\bar z_{12}^{2h_w}\bar z_{34}^{2h_v}}
	\underbrace{\big[z^{2\alpha_1\alpha_2+2h_w}(1-z)^{2\alpha_2\alpha_3}\times c.c. \big] G(z,\bar z)}_{f_{WV}(z,\bar z)}.
\end{align}
The function $G(z,\bar z)$ is a sum of two independent functions $I_1, I_2$ corresponding to two fusion channels respectively,
\begin{align}
	G(z,\bar z) = A\left[\frac{s(b)s(a+b+c)}{s(a+c)} |I_1(a,b,c,z)|^2 + \frac{s(a)s(c)}{s(a+c)} |I_2(a,b,c,z)|^2 \right],
\end{align}
where $a=2\alpha_-\alpha_1$, $b=2\alpha_-\alpha_3$, $c=2\alpha_-\alpha_2=a$, $s(a) = \sin a\pi$ and $A$ is a normalization factor. One can fix it by the operator algebra, but it is irrelevant to our discussion here. Moreover, we use the notation $|h(z)|^2=h(z)h(\bar z)$ to simplify our formula. 

$I_1, I_2$ are related to Hypergeometric functions respectively,
\begin{align}\nonumber
	I_1(a,b,c,z) = & \frac{\Gamma(-a-b-c-1)\Gamma(b+1)}{\Gamma(-a-c)} F(-c,-a-b-c-1;-a-c;z), \\ \nonumber
	I_2(a,b,c,z) = & z^{1+a+c} \frac{\Gamma(a+1)\Gamma(c+1)}{\Gamma(a+c+2)} F(-b,a+1;a+c+2;z).
\end{align}
One can do a small $z,\bar z$ expansion of the above expression to see that $I_1$ comes from the fusion channel $[\phi_{1,3}]$ and $I_2$ comes from $[\phi_{1,1}]$. If $z$ winds clockwise around $z=1$, then $(I_1, I_2)^T$ will transforms under the following monodromy matrix,
\begin{align}
	M = 
	\frac{1}{s(a+c)s(b+c)} 
	\left(
	\begin{array}{cc}
		s(a)s(b)+\lambda s(c)s(a+b+c) & (-1+\lambda)s(a)s(c) \\
		(-1+\lambda)s(b)s(a+b+c) & \lambda s(a)s(b) + s(c)s(a+b+c)
	\end{array}
	\right),
\end{align}
where $\lambda = e^{-i2\pi(b+c)}$.

\subsection{Normal-order correlation functions}

\begin{figure}[!h]
	\centering
	\includegraphics[width=0.5\textwidth]{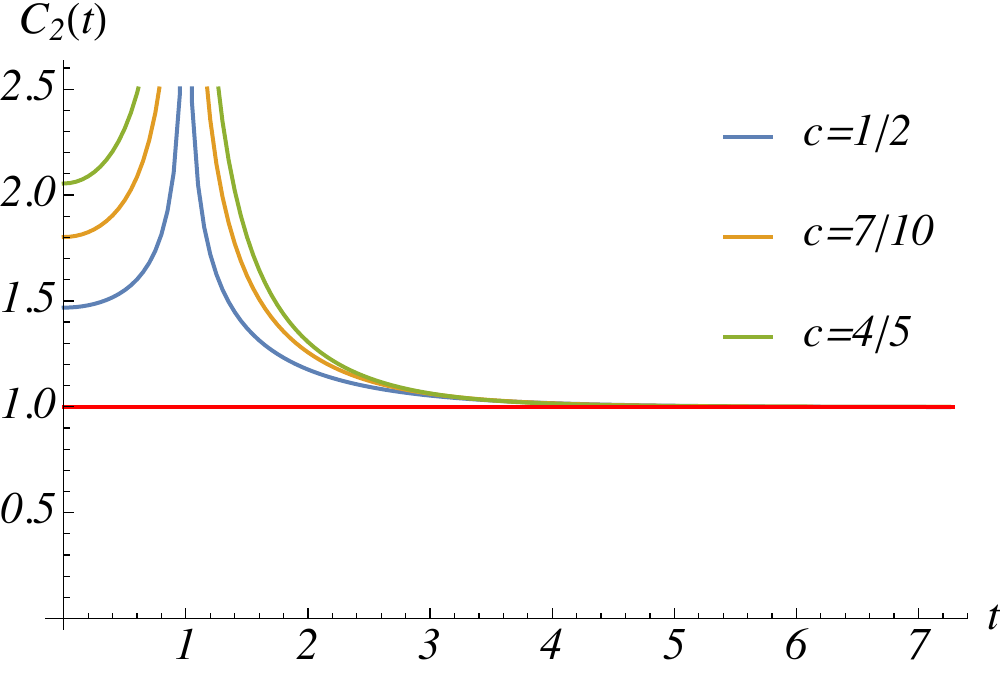}
	\caption{\label{fig:noc} Normal-order correlator for $W=V=\phi_{1,2}$ in different minimal models. We choose $x=1$, $\beta=2\pi$ and $\epsilon=0.0001$. The red line is constant $1$, plotted for convenience. $C_2(t)$ diverges at $t=x$, which is the light cone singularity. Because we drop the prefactor, $C_2(t)$ approaches $1$, i.e., the equilibrium value.}
\end{figure}

Let us first discuss the normal-order correlator. As an example, we choose $M(4,3)$ and $W=V=\phi_{1,2}$. The result is shown by the blue curve in Fig.~\ref{fig:noc}. (Here we drop the prefactor $1/z_{12}^{2h_w}z_{34}^{2h_v}$ and only plot $f_{WV}$, same for all the results below) We can see that it starts from some non-generic value depending on $x/\beta$ and operator content. Then it diverges at the light cone $t=x$ because the two operators hit each other. Then at a time scale $\beta$, the excitations at $W$'s position created by $V$ collide and decay into equilibrium and the NOC approaches its final value $\braket{WW}\braket{VV}$.

As a comparison, we fix the choice of operator and tune the central charge. The result is shown in Fig.~\ref{fig:noc}. We can see that different curves share similar behaviors. Physically, one may expect the NOC for larger central charge to decay faster because there are more degrees of freedom for larger central charge and there will be more collision to get into equilibrium. This is indeed what the results in Fig.~\ref{fig:noc} show. However, if we take a more technical viewpoint, we will find that this is because the late time decay rate is controlled by the intermediate channels of the $WW$ OPE, which is $W\times W=\phi_{1,1} \times \phi_{1,3}$, and $h_{1,3}$ increases with central charge. If we choose different operators, for example, using $W=V=\phi_{2,3}$ then $\phi_{3,3}$ will appear in the fusion channel and has the second smallest conformal weight. But $h_{3,3}$ will decrease with central charge. Hence, larger central charge doesn't necessarily mean faster equilibrium and this behavior actually depends on the choice of operator.

\begin{figure}[!t]
	\centering
	\includegraphics[width=0.9\textwidth]{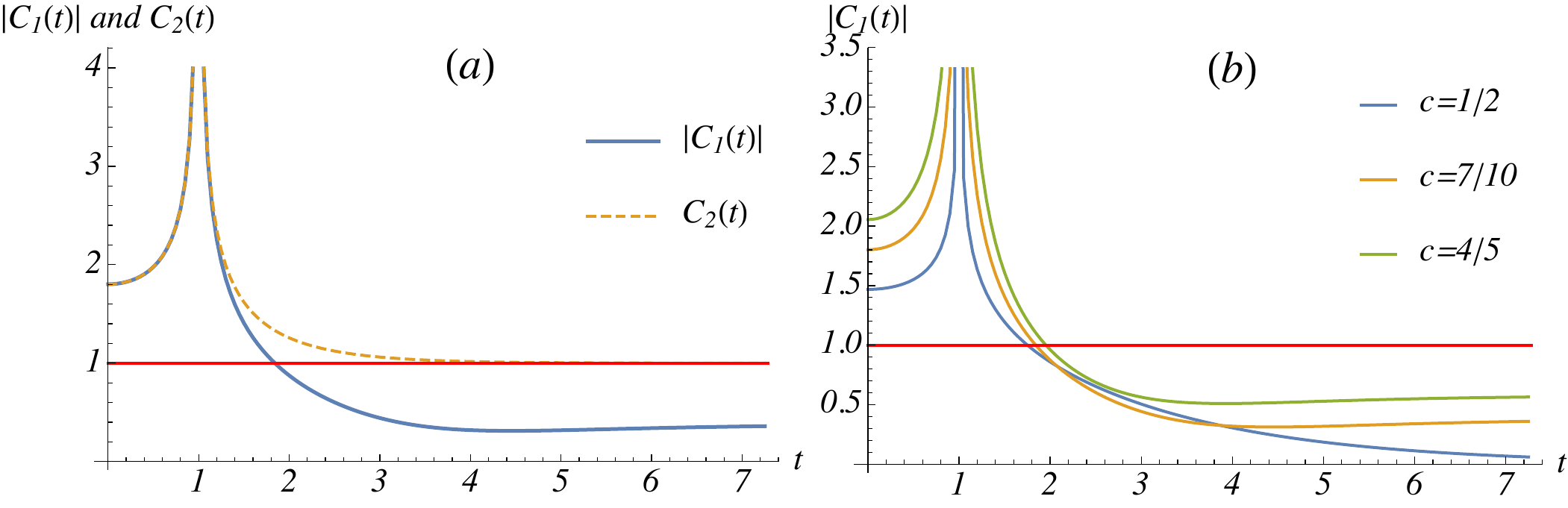}
	\caption{\label{fig:otoc} OTOCs and NOC for $W=V=\phi_{1,2}$. (a) OTOC for $M(5,4)$. (b) OTOC for different different models labeled by their central charges. We choose $x=1$, $\beta=2\pi$ and $\epsilon=0.0001$. The red line is constant $1$, plotted for convenience.}
\end{figure}

\subsection{Out-of-time-order correlation functions}

We now discuss the OTOC. We choose $M(5,4)$ and $W=V=\phi_{1,2}$ as an example. The result is shown in Fig.~\ref{fig:otoc}(a), which is the absolute value of OTOC. We also plot the NOC for comparison. 

We can see that there is no difference between OTOC and NOC at the beginning. Both have a singularity at the light cone $t=x$. Their differences only appear at later times. The NOC measures local equilibrium processes, so it approaches $\braket{W W}_\beta\braket{V V}_\beta$ at late times. The OTOC measures scrambling, so it continues to decay to a smaller value. As we can read from the figure, both happen at the same time scale $\beta$ after $W$ and $V$ hit each other, so this confirms our assertion of no separation of time scales. 

In Fig.~\ref{fig:otoc}(b), we show the result for different unitary minimal models. We can see that the late-time behavior highly depends on the model and there is a tendency that the final value gets closer to 1 as central charge increases. This is because as $c\rightarrow 1$, the system will approach a free boson, where we have well-defined quasi-particles and we don't expect scrambling.

\subsection{Early-time and late-time behaviors}

Because of the light cone singularity, it is hard to directly consider the early-time behavior of the OTOC and NOC. Also, the late-time value of OTOC itself is very not meaningful. Only its ratio with NOC diagnoses scrambling. So in this section, we will consider the ratio between the absolute value of the OTOC and NOC, $f(t)=|C_1(t)|/C_2(t)$.

\begin{figure}[!h]
	\centering
	\includegraphics[width=0.6\textwidth]{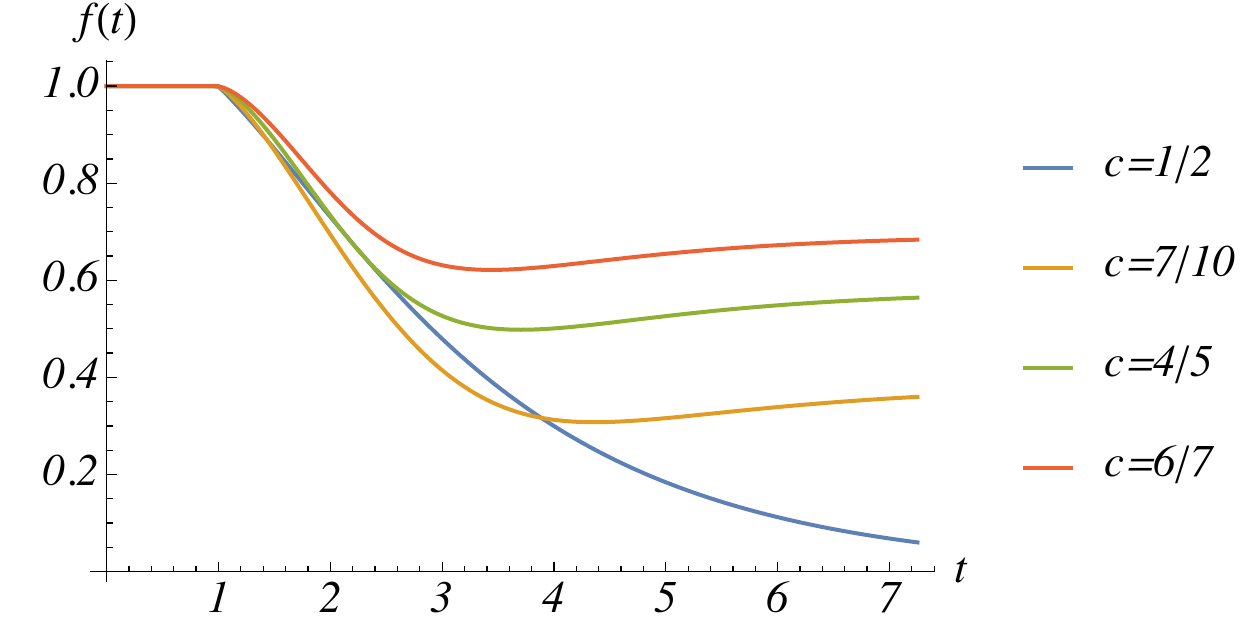}
	\caption{\label{fig:ratio} Ratio between OTOC and NOC for $W=V=\phi_{1,2}$ and different central charges. We choose $x=1$, $\beta=2\pi$ and $\epsilon=0.0001$.}
\end{figure}

We choose $W=V=\phi_{1,2}$ and do the calculation in different models. The result is plotted in Fig.~\ref{fig:ratio}. We can clearly see that $f(t)=1$ before $V$ enters the light cone of $W$. When $t>x$, both thermalization and scrambling happen at the same speed. So $f(t)$ gets to a final value at the time scale of order $\beta$. When the central charge is larger, the final value of $f(t)$ is closer to $1$, which indicates that the scrambling becomes weaker. 

\begin{figure}[!h]
	\centering
	\includegraphics[width=0.9\textwidth]{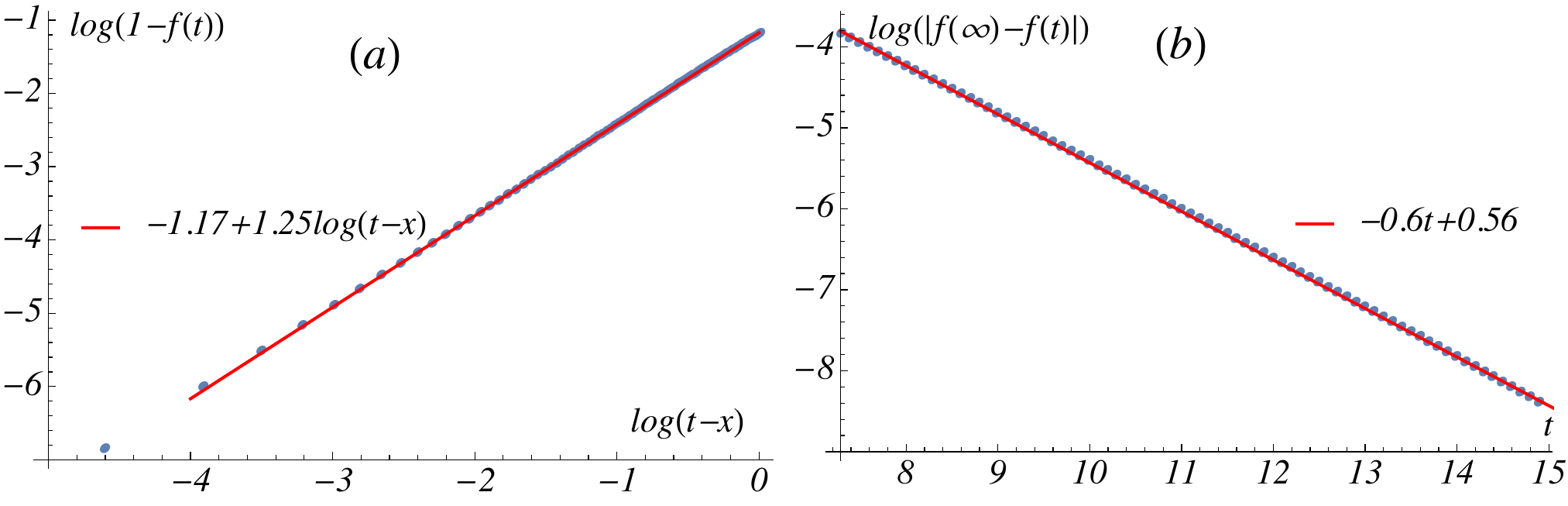}
	\caption{\label{fig:fitting} Fitting of the early-time and late-time behavior of the ratio $f(t)$. (a) Early-time behavior. The slope of the fitting curve is $1.25$, which is close to $2(h_{1,3}-h_{1,1})=1.2$. (b) Late-time behavior. The slope of the fitting curve is $0.60$, which is close to $h_{1,3}=0.6$. We choose $x=1$, $\beta=2\pi$ and $\epsilon=0.0001$.}
\end{figure}

In Sec.\ref{sec:argument}, we have argued that the early-time behavior of $f(t)$ follows $1+ \# (t-x)^r$. and the late-time behavior is $|M_{p_1p_1}| + \# \exp\left(-s\frac{2\pi}{\beta}t \right)$. Now that we know the analytical result, we can use this data to do an explicit check. Here we just show a typical example where $W=V=\phi_{1,2}$ and $c=7/10$. Both the $WW$ and $WV$ OPE are $\phi_{1,2}\times\phi_{1,2}=\phi_{1,1}+\phi_{1,3}$. So from our argument, we expect an exponent $r=2(h_{1,3}-h_{1,1})=6/5$ for the early-time decaying and an exponent $s=h_{1,3}=3/5$ for the late-time behavior. The fitting result is shown in Fig.~\ref{fig:fitting}, which is consistent with our analysis.

\section{Conclusion}
\label{sec:conclusion}

In this paper, we analyzed the OTOC $C_1(t)$ and the NOC $C_2(t)$ of primary fields $W$ and $V$ in unitary minimal models with infinite system size and finite temperature $1/\beta$.

Qualitatively, based on the general structure of the four point function and the second-sheet effect, we can conclude the following picture. We put $V$ at $x$ and evolve $W$ at the origin. In CFTs, the conformal symmetry guarantees that $W(t)$ expands with a sharp light cone. When $t<x$, $V$ hasn't entered the light cone of $W$ thus both scrambling and thermalization haven't started yet. Thus, we have $C_1(t)=C_2(t)$ in this time regime. At $t=x$, $V$ hits the front of $W(t)$ and both $C_1(t)$ and $C_2(t)$ develop a light cone singularity here. When $t>x$, the NOC approaches the equilibrium value $\braket{WW}_\beta\braket{VV}_\beta$ while the OTOC decays to a smaller value controlled by monodromy. Because of the lack of large parameter, there is no separation of time scales so these two processes happen at the same time scale $t-x\sim\beta$. 

More quantitatively, we can further divide $t>x$ into two time regimes and study the ratio $f(t)=C_1(t)/C_2(t)$. At early times $0<t-x\ll\beta$, $f(t)$ decays as $e^{2\pi i (h_{q_1}-h_W-h_V)}(1+\# (t-x)^r)$ with a fractional power $r=2(h_{q_2}-h_{q_1})>0$, where $h_{q_2}$and $h_{q_1}$ are the two smallest conformal weights of $WV$ fusion channels. At late times, $f(t)$ will exponentially approach a late-time value as $M_{p_1p_1}+\#\exp(-s\frac{2\pi t}{\beta})$, where $p_1$ is the identity channel and $s=h_{p_2}$ is the second smallest conformal weight of $WW$ and $VV$ fusion channels.

Furthermore, all of these pictures can be confirmed by an explicit calculation using the Coulomb gas formalism. With these analytical results, we can also tune the central charge at will. As $c$ increases from $1/2$ to $1$, on the one hand the equilibration and scrambling processes do not necessarily become faster but depend on the choice of operators. On the other hand, $f(t)$ becomes larger indicating weaker scrambling, which is also consistent with our understanding that under the Coulomb gas formalism, the system becomes a free boson when $c\rightarrow 1$ and thus shouldn't have scrambling.

Previously, both early-time and late-time behaviors have been calculated in various models. However, it hasn't been well understood whether and how they are related. This paper shows a connection in the context of minimal models. So it would be interesting to think about more implications of such connections and its generalization to other models.

There is a lot of recent progress on studying scrambling in (1+1)D systems including random circuit and some lattice models. It was found that for two spatially separated local operator $W$ and $V$, $\braket{|[W(t),V(0)]|^2} \sim \exp\left( -\# (t-x)^{p+1}/t^p \right)$ at early time, with $0\le p\le 1$~\cite{lattice5, lattice6}. Unitary minimal models can be thought of as the IR limit of those 1D lattice models. Due to the intrinsic divergence of field theory, we cannot study the same quantity. However, we still observe a fractional power $r$ for the early-time regime. It will be interesting to connect these two results.

\section*{Acknowledgments}

We would like to thank Ashvin Vishwanath and Shang Liu for stimulating discussions and Nathanan Tantivasadakarn and Pengfei Zhang for helpful comments on this manuscript. We especially thank Yingfei Gu for many important discussions on both the physical picture and technical details of this work. We also would like to thank helpful questions and comments during the seminar at Tsinghua University, Institute for Advanced Study. We acknowledge funding from  the Simons Foundation through Ashvin Vishwanath's Simons Investigator grant. R. F. was also supported by the Purcell fellowship when this work was done.

\appendix

\section{The Coulomb gas formalism}
\label{app:coulomb}

In this appendix, we will give a brief introduction to the Coulomb gas formalism\cite{coulombgas}. This method uses 2D free boson with a special boundary condition. The primary fields of minimal models are built from the vertex operators of boson field. Any correlation function of primary fields can be translated into a corresponding correlation function of vertex operators, which is easier to compute. 

We will start from free boson and sketch the construction of screened Coulomb gas model. We will see that after selecting a certain class of vertex operators and constraining the value of the central charge, the screened Coulomb gas will have the same central charge, primary field contents and fusion rules as minimal models. That's how we make the identification between them. For a more rigorous treatment, readers can refer to \cite{brst} or the corresponding chapter of \cite{yellowbook}. Finally, we will show how to use this formalism to compute the parent four point function that we use in the main text.

\subsection{Free boson}

The free boson CFT can be described by a path integral with the following action,
\begin{align}
	S = \frac{1}{8\pi} \int d^2 x \partial_\mu \varphi \partial^\mu \varphi 
	= \frac{1}{8\pi} \int d^2 x \left( \partial_z \varphi \partial^z \varphi + \partial_{\bar z} \varphi \partial^{\bar z} \varphi \right). 
\end{align}
The two point function $K({z,\bar z}) = \braket{\varphi(z,\bar z)\varphi(0)}$ can be solved from the equation of motion,
\begin{align}
	\partial_{\bf x}^2 K({\bf x}) = -4\pi \delta^{(2)}({\bf x}) \Rightarrow
	K(z,\bar z) = -\log |z\bar z|.
\end{align}
This functional form is the same as the 2D Coulomb interaction. That's why the name Coulomb gas is used.

Conformal symmetry yields a chiral stress-energy tensor,
\begin{align}
	T = -2\pi T_{zz} = -\frac{1}{2} \partial_z \varphi \partial_z \varphi,\quad 
	\bar T = -2\pi T_{\bar z\bar z} = -\frac{1}{2} \partial_{\bar z} \varphi \partial_{\bar z} \varphi.
\end{align}
Using the two point function, one can show the associated Virasoro algebra has central charge $c=1$. Primary fields are $\partial_z\varphi(z,\bar z)$ with conformal weight $h=1,\bar h=0$ and the vertex operator $V_\alpha(z,\bar z)=:e^{i\sqrt{2}\alpha\varphi(z,\bar z)}:$ with conformal weight $h=\bar h=\alpha^2$.

This system also has a $U(1)$ symmetry $\varphi \rightarrow \varphi + a$ with $a$ to be a real constant. This symmetry gives us another current,
\begin{align}
	J(z) = \frac{i}{2}\partial_z \varphi(z,\bar z),\quad
	\bar J(\bar z) = -\frac{i}{2} \partial_{\bar z} \varphi(z,\bar z).
\end{align}
So the Virasoro algebra is augmented with a $U(1)$ Kac-Moody algebra. If we require our primary fields are primary with respect to both algebras, then we are left with vertex operators $V_\alpha(z,\bar z)$ which have charge $\sqrt{2}\alpha$.\footnote{
	We will drop the unimportant $\sqrt{2}$ factor when talking about the charges of vertex operators below.
} Their correlation functions are,
\begin{align}
	\braket{V_{\alpha_1}(z_1, \bar z_1) V_{\alpha_2}(z_2, \bar z_2)...V_{\alpha_n}(z_n, \bar z_n)} = \delta_{\sum_i \alpha_i,0}\prod_{i<j} |z_i-z_j|^{4\alpha_i \alpha_j}.
\end{align}
The constraint $\sum_i\alpha_i = 0$ is called neutrality condition and comes from the $U(1)$ symmetry. We can interpret this correlation function as a partition function of a charge neutral Coulomb gas. All particles sit at $z_1,z_2,...,z_n$ with Coulomb interaction between them.
 
\subsection{Screened Coulomb gas}

Now we modify the boundary condition by adding a background charge $-2\alpha_0<0$ at the infinity. Then the neutrality condition for vertex operator correlators becomes $\sum_i \alpha_i = 2\alpha_0$. This modification has two important effects:
\begin{itemize}
	\item The conformal weight of the vertex operator is changed to
	\begin{align}
		h_\alpha = h_{2\alpha_0-\alpha} = \alpha^2 - 2\alpha_0\alpha.
	\end{align}
	\item To be self-consistent, the stress-energy tensor also has to be modified correspondingly,
	\begin{align}\label{eqn:t_screen}
		T = -\frac{1}{2}:\partial\varphi\partial\varphi: + i\sqrt{2}\alpha_0:\partial^2 \varphi:.
	\end{align} 
	As a result, the central charge becomes smaller $c=1-24\alpha_0^2$.\footnote{
		Instead of adding the background charge by hand, another approach is to modify the free boson action by coupling it to the Ricci curvature, $S = \frac{1}{8\pi}\int dx^2 \sqrt{g}(\partial_\mu \varphi \partial^\mu \varphi + i2\sqrt{2}\alpha_0 \varphi R)$. We consider the geometry that the space is everywhere flat so $\braket{\varphi(z,\bar z)\varphi(w,\bar w)}$ doesn't change. But we require it has the same topology as a sphere. Thus, $R=0$ almost everywhere except at the infinity to satisfy the Gauss-Bonnet theorem $\int dx^2 R = 8\pi$. One can show this action gives the same stress-energy tensor and neutrality condition for vertex operator correlators as we write above.
	}
\end{itemize}

Then, we want to use this framework to describe a physical system at criticality, i.e. identify physical observables having definite scaling dimension with vertex operator having the same scaling dimension. Naturally, $V_\alpha$ and $V_{2\alpha_0-\alpha}$ have the same conformal dimension thus should correspond to the same physical operator $\phi_\alpha$. Thus, it's tempting to write down $\braket{\phi_\alpha \phi_\alpha } ``=" \braket{V_\alpha V_\alpha} ``=" \braket{V_\alpha V_{2\alpha_0-\alpha}}$. However, only the second one is nonzero. For a four point function $\braket{\phi_\alpha\phi_\alpha\phi_\alpha\phi_\alpha}$, we even cannot write down a nonzero result if we want to identify it to vertex operator. The identification doesn't work in the current setting. The way out is to introduce screening operators.

As the term suggests, this operator when inserted into correlation function can screen some charge to help satisfy the neutrality condition. So it must carry charge, i.e. composed of vertex operators. On the other hand, we don't want to change the conformal properties of the correlator, so the conformal dimension of screening operators must be zero. Thus the simplest choice is 
\begin{align}
	Q_{\pm} = \oint dz V_{\alpha_\pm}(z),\quad \alpha_{\pm} = \alpha_0 \pm \sqrt{\alpha_0^2 + 1}
\end{align}
$\alpha_{\pm}$ is chosen in order for $h_{\alpha_{\pm}} = \alpha_\pm^2 - 2\alpha_\pm \alpha_0 = 1$. The following two formula of $\alpha_{\pm}$ are more useful for later usage,
\begin{align}\label{eqn:alphapm}
	\alpha_+ + \alpha_- = 2\alpha_0, \quad 
	\alpha_+ \alpha_- = -1.
\end{align}
The contour in the definition of $Q_\pm$ is not fixed but determined by other operators in the correlator and some physical requirements. We will see examples in the next section.

Now given the background charge, we can classify the vertex operators into two categories:
\begin{itemize}
	\item {\bf physical} For a vertex operator $V_\alpha$ (and its dual $V_{2\alpha_0-\alpha}$), if we can choose an appropriate number of screening operators such that,
	\begin{align}\label{eqn:physical_requirement}
		\braket{V_\alpha V_{\alpha} Q_+^r Q_-^s} = \braket{V_{\alpha}V_{2\alpha_0-\alpha}}
	\end{align}
	then we can identify it with a physical observable. This requirement gives a constraint on $\alpha$ via the neutrality condition,
	\begin{align}\nonumber
		2 \alpha + r \alpha_+ + s\alpha_- = 2\alpha_0 = \alpha_+ + \alpha_-. \\
		\Rightarrow \alpha_{r,s} = \frac{1}{2}(1-r)\alpha_+ + \frac{1}{2}(1-s)\alpha_-.
	\end{align}
	So each physical vertex operator is labeled by  an index $(r,s)$ and its dual $V_{2\alpha_0-\alpha}$ has an index $(-r,-s)$. Both of them have the conformal weight,
	\begin{align}
		h_{r,s} = \frac{1}{4}(r\alpha_+ + s\alpha_-)^2 - \alpha_0^2.
	\end{align}
	One can show that the operators satisfying these constraints also yield nonzero higher order correlator with itself. We will identify these kind of vertex operators with physical operators $\phi_{r,s}$ with the same conformal weight. So the physical correlation function can be converted to a correlator of physical vertex operators with screening operators, which is a integral of an already know function thus can be worked out in principle.
	\item {\bf unphysical} Those vertex operators that don't satisfy this requirement Eqn\eqref{eqn:physical_requirement} are unphysical. In the following, we only consider the physical vertex operator subset.
\end{itemize}

\subsection{Identification with minimal models: fusion rules}

Although the physical requirement has already selected a small portion of operators, we still have infinite number of primary fields. One way to truncate the operators is to require 
\begin{align}\label{eqn:rational_requirement}
	p'\alpha_+ + p\alpha_- = 0
\end{align}
which combined with Eqn\eqref{eqn:alphapm} fixes the central charge and conformal weight of vertex operators to be,
\begin{align}
	c & = 1-\frac{6(p-p')^2}{pp'}, \\ \label{eqn:hrs}
	h_{r,s} & = \frac{(rp-sp')^2 - (p-p')^2}{4pp'}.
\end{align}
These formula look exactly the same as those of minimal models except that the two integers $r,s$ are unbounded. To demonstrate this model represents minimal model, we need to show the values of $r,s$ are restricted by $p,p'$. 

Such kind of restriction does emerge if we look at the fusion rules, i.e. by studying the three point function of physical operators $\braket{\phi_{k,l}\phi_{m,n}\phi_{r,s}}$. After mapping it to correlation function of vertex operators and doing the integral\cite{brst}, we will see it vanishes unless,
\begin{align} \nn
	|m-r|+1 \le k \le \min(m+r-1,2p'-m-n-1) \\
	|n-s|+1 \le l \le \min(n+s-1,2p-n-s-1).
\end{align}
Only if we require $1\le r< p', 1\le s<p$ in Eqn\eqref{eqn:hrs} can we satisfy the fusion rules above.

Up to now we demonstrate that in the screened Coulomb gas with constraint Eqn\eqref{eqn:rational_requirement}, vertex operators $V_{r,s}$ with conformal weight $h_{r,s}=\bar h_{r,s}$ form a closed subalgebra, which can be identified with minimal models. In the main text, we focus on the unitary minimal models.

\subsection{Examples of computing four point correlators} 
Now let's consider 4pt functions of primary fields. Unlike the two point and three point functions, conformal invariance can only restricts the four point functions to the following form,
\begin{align}
	\braket{\phi_1(z_1,\bar z_1) \phi_2(z_2,\bar z_2) \phi_3(z_3,\bar z_3) \phi(z_4,\bar z_4)} & = f(\eta,\bar \eta) \prod_{i<j} z_{ij}^{\mu_{ij}} \bar z_{ij}^{\bar \mu_{ij}}, \\
	\eta = \frac{z_{12}z_{34}}{z_{13}z_{24}},\quad
	\mu_{ij} = \frac{1}{3}(\sum_{k=1}^4 h_k) & - h_i - h_j.
\end{align}
$f(\eta,\bar \eta)$ can only fixed by dynamical data.

In minimal models $f(\eta, \bar \eta)$ can be calculated by solving some differential equations. Now the Coulomb gas formalism provides another approach. Under this scheme, we can use the identification between primaries and vertex operators to rewrite the 4pt function of physical observables as a correlation function of vertex operators with some screening operators,
\begin{align} \nonumber
	& \braket{\phi_1(z_1,\bar z_1) \phi_2(z_2,\bar z_2) \phi_3(z_3,\bar z_3) \phi(z_4,\bar z_4)} \\
	= &
	\braket{V_{\alpha_1}(z_1,\bar z_1) V_{\alpha_2}(z_2,\bar z_2) V_{\alpha_3}(z_3,\bar z_3) V_{\alpha_4}(z_4,\bar z_4)Q_+^m Q_-^n \bar Q_+^{\bar m} \bar Q_-^{\bar n}},
\end{align}
so that the R.H.S gives us an integral formula. By evaluating the integral and comparing the result with the L.H.S, we can determine $f(\eta,\bar \eta)$ and the full four point function.

As an concrete example, let's consider the parent four point correlator that we used in the main text,
\begin{align}\label{eqn:2nd_4pt}
	\braket{\phi_{(1,2)}\phi_{(1,2)}\phi_{(m,n)}\phi_{(m,n)}}=\braket{V_{1,2}V_{1,2}V_{m,n}V_{-m,-n}Q_-\bar Q_-}.
\end{align} 
For $V_{(2,1)}$, we need to insert $Q_+$ instead. In the following, we label the four operators by $1,2,3,4$ from left to the right to make our formula cleaner. For example, $z_1,\alpha_1$ representing the coordinate and charge of the first operator and so on.

\begin{figure}[t]
	\centering
	\includegraphics[width=0.8\textwidth]{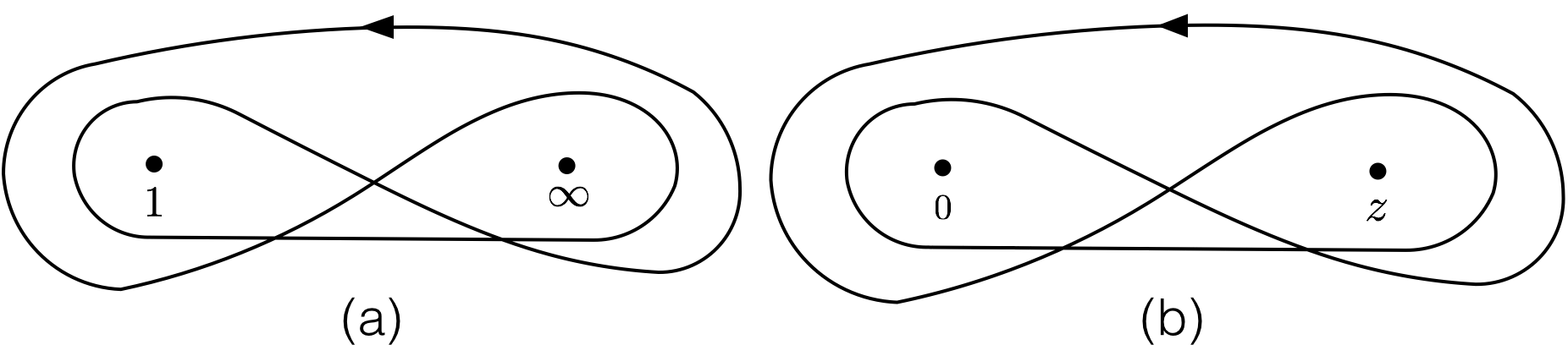}
	\caption{\label{fig:pochhammer} Two independent choices of integration contour.}
\end{figure}

The integrals in $Q_-$ and $\bar Q_-$ are independent of each other. Let's first look at its holomorphic part,
\begin{align}
	\braket{V_{1,2}V_{1,2}V_{m,n}V_{-m,-n}Q_-}=
	\prod_{k<j}^4 z_{kj}^{2\alpha_k\alpha_j} \oint d\omega \prod_{j=1}^4(\omega - z_j)^{2\alpha_+\alpha_j}.
\end{align}
To simplify the calculation, we take a conformal transformation so that $z_1 \rightarrow 0$, $z_2 \rightarrow z$, $z_3 \rightarrow 1$, $z_4 \rightarrow \infty$, $\eta \rightarrow z$. And we have,
\begin{align} \nonumber
	& \lim_{z_4\rightarrow \infty} z_4^{2h_4} \braket{V_{1,2}(0)V_{1,2}(z)V_{m,n}(1)V_{-m,-n}(z_4)Q_-} \\
	= & z^{2\alpha_1\alpha_2}(1-z)^{2\alpha_2\alpha_3} \oint d\omega\, \omega^a (\omega-1)^b (\omega-z)^c \\
	= & z^{2\alpha_1\alpha_2}(1-z)^{2\alpha_2\alpha_3} I(a, b, c, z)
\end{align}
where $a=2\alpha_-\alpha_1$, $b=2\alpha_-\alpha_3$, $c=2\alpha_-\alpha_2$. The integrand has four branching points $0,z,1,\infty$. If we choose an arbitrary contour, the result will depend on the choice of branch cut. However, the physics should be insensitive to the position of branch cut. To eliminate this unphysical sensitivity, we choose Pochhammer double contour, which guarantees that each branch cut is crossed twice in opposite direction so that we go back to the original Riemann surface. There are two independent choices shown in Fig.~\ref{fig:pochhammer}. 
We can write down two independent solutions for the integration:
\begin{align}\nonumber
	I_1(a,b,c,z) = & \int_P^{(1,\infty)} d\omega\, \omega^a (\omega-1)^b (\omega-z)^c \\ \nonumber
	\overset{\omega=1/t}{=} &
	\int_P^{(0,1)} dt\, t^{-a-b-c-2}(1-t)^{b}(1-zt)^c \\
	= & \frac{\Gamma(-a-b-c-1)\Gamma(b+1)}{\Gamma(-a-c)} F(-c,-a-b-c-1;-a-c;z), \\ \nonumber
	I_2(a,b,c,z) = & \int_P^{(0,z)} d\omega\, \omega^a (\omega-1)^b (\omega-z)^c \\ \nonumber
	\overset{\omega=zt}{=} & z^{1+a+c} \int_P^{(0,1)} dt\, t^a (1-t)^c (1-zt)^b \\ 
	= & z^{1+a+c} \frac{\Gamma(a+1)\Gamma(c+1)}{\Gamma(a+c+2)} F(-b,a+1;a+c+2;z).
\end{align}
where $\int_P$ means integration along the Pochhammer contour. In the final step of the two equations, we use the integral representation of Hypergeometric function (see Appendix \ref{app:hypergeo}). Here we're sloppy with the unimportant phase factors and drop the prefactors in the two final results\footnote{
	The final coefficients will be determined by monodromy invariance. Therefore, it is not necessary to include all the prefactors here. The only useful information contained in the dropped prefactors is that $a,b,c$ cannot be integers otherwise the integral will vanish.
}. 

This Pochhammer contour integration can be simplified when $a>-1, b>-1, c>-1, a+b+c<-1$. We can always deform the contour to be composed of several lines connecting the branching point and small circles around the branching points. For this parameter regime, the integral along small circles goes to zero if the radius of circles goes to zero. Then we can get rid of the circle integrals and only keep those line integrals. If we don't care about the unimportant phase factor, we can reduce the contour integral to the following line integral,
\begin{align}
	I_1(a,b,c,z) = & \int_1^\infty d\omega\, \omega^a (\omega-1)^b (\omega-z)^c,\quad \text{when } b>-1, a+b+c<-1 ,\\
	I_2(a,b,c,z) = & \int_0^z d\omega\, \omega^a (\omega-1)^b (\omega-z)^c,\quad \text{when } a>-1, c>-1.
\end{align}
This representation will be useful when we discuss the monodromy problem.

The consideration of the anti-holomorphic part will be completely the same except for replacing $w,z$ with $\bar w, \bar z$. So the anti-holomorphic part also have two independent solutions $I_1(a,b,c,\bar z)$ and $I_2(a,b,c,\bar z)$. The physical four point function will be a linear combination of them and thus is,
\begin{align}
	|z|^{4\alpha_1\alpha2} |1-z|^{4\alpha_2\alpha_3} \sum_{ij}X_{ij}I_i(z) \overline{I_j(z)} = 
	|z|^{4\alpha_1\alpha2} |1-z|^{4\alpha_2\alpha_3} G(z,\bar z)
\end{align}
The coefficients $X_{ij}$ can be completely determined by monodromy invariance and conformal algebra.

\paragraph{Monodromy Invariance} 

The conformal blocks have branch cut on the complex plane. Therefore, if we move the cross ratio around those singular points by $2\pi$, conformal blocks may not go back to its original value but pick up a monodromy matrix, i.e. $F_p \rightarrow \sum_{p'}M_{pp'}F_{p'}$. However, locality condition tells us the physical correlator should be invariant under such an operation. This is called monodromy invariance, which gives us constraints on $X_{ij}$. Here it is enough to examine the monodromy around $z=0$ and $z=1$.

At $z=0$, $F(\alpha,\beta,\gamma,z)$ has a Taylor expansion. So if we drag $z$ around $z=0$, only the $z^{1+a+c}$ factor in $I_2$ gives a non-trivial phase factor. The corresponding monodromy matrix is,
\begin{align}
	M_{z=0} = \left(  
	\begin{array}{cc}
		1 & 0 \\
		0 & e^{2\pi i(1+a+c)}
	\end{array}
	\right).
\end{align}
This requires $X_{ij}$ to be diagonal and we simply write it as $X_j$.

To study the monodromy invariance around $z=1$, it's easier to first express $I_j(z)$ in terms of $I_j(1-z)$, which is a linear relation $I_i(z) = \sum_j a_{ij} I_j(1-z)$. Then we have $G(z,\bar z)=\sum_{ijk} X_ia_{ij}a_{ik}I_j(1-z)\bar{I_k(1-z)}$. The monodromy matrix for $I_j(1-z)$ around $z=1$ is also a diagonal matrix of some phase factor. So the invariance requires $\sum_i X_ia_{ij}a_{ik}$ to be diagonal with respect to $j,k$, which will finally fix $X_1/X_2$.

\begin{figure}[t]
	\centering
	\begin{tikzpicture}[>=latex]
	
	\def\ra{0.4}
	\def\ya{0}
	\def\yb{-1.3}
	\def\yc{-2.6}
	\def\yd{-3.9}
	\def\ye{-5.2}
	\def\xstart{0}
	\def\xend{8}
	\def\xa{2}
	\def\xb{4}
	\def\xc{6}
	\def\offset{0.3}
	
	\draw (\xstart,\ya) -- (\xend,\ya);
	\foreach \Point in {(\xa,\ya), (\xb,\ya), (\xc,\ya)}{
		\draw[fill=black] \Point circle [radius=1.2pt];
	}
	\draw[semithick, blue,decoration={markings, mark=at position 0.5 with 		{\arrow{>}}}, postaction={decorate}] (\xc,\ya) -- (\xend+0.3,\ya);
	\node at (\xend+0.3,\ya) [anchor = west] {$\times (e^{i\pi(b+c)}-e^{-i\pi(b+c)})$};
	
	\draw[decoration={brace,mirror,raise=5pt},decorate]
	(-0.2,\yb) -- node[left=6pt] {$=$} (-0.2,\yc);
	
	\node at (\xstart,\yb) {$-$};
	\draw (\offset+\xstart,\yb) -- (\offset+\xend,\yb);
	\foreach \Point in {(\offset+\xa,\yb), (\offset+\xb,\yb), (\offset+\xc,\yb)}{
		\draw[fill=black] \Point circle [radius=1.2pt];
	}
	\draw[semithick, blue,decoration={markings, mark=at position 0.5 with 		{\arrow{>}}}, postaction={decorate}] (\offset+\xc,\yb) -- node[anchor = south] {\color{black} $e^{i\pi b}$} (\offset+\xb+\ra,\yb);
	\draw[semithick, blue,decoration={markings, mark=at position 0.5 with 		{\arrow{>}}}, postaction={decorate}] (\offset+\xb-\ra,\yb) -- node[anchor = south] {\color{black} $e^{i\pi(b+c)}$} (\offset+\xa+\ra,\yb);
	\draw[semithick, blue,decoration={markings, mark=at position 0.5 with 		{\arrow{>}}}, postaction={decorate}] (\offset+\xa-\ra,\yb) -- node[anchor = south] {\color{black} $e^{i\pi(a+b+c)}$} (\offset+\xstart,\yb);
	\draw[semithick, blue] (\offset+\xb+\ra,\yb) arc [radius=\ra, start angle=0, end angle=180];
	\draw[semithick, blue] (\offset+\xa+\ra,\yb) arc [radius=\ra, start angle=0, end angle=180];
	\node at (\offset+\xend,\yb) [anchor=west] {$\times e^{-i\pi(b+c)}$};
	
	\node at (\xstart,\yc) {$+$};
	\draw (\offset+\xstart,\yc) -- (\offset+\xend,\yc);
	\foreach \Point in {(\offset+\xa,\yc), (\offset+\xb,\yc), (\offset+\xc,\yc)}{
		\draw[fill=black] \Point circle [radius=1.2pt];
	}
	\draw[semithick, blue,decoration={markings, mark=at position 0.5 with 		{\arrow{>}}}, postaction={decorate}] (\offset+\xc,\yc) -- node[anchor = south] {\color{black} $e^{-i\pi b}$} (\offset+\xb+\ra,\yc);
	\draw[semithick, blue,decoration={markings, mark=at position 0.5 with 		{\arrow{>}}}, postaction={decorate}] (\offset+\xb-\ra,\yc) -- node[anchor = south] {\color{black} $e^{-i\pi(b+c)}$} (\offset+\xa+\ra,\yc);
	\draw[semithick, blue,decoration={markings, mark=at position 0.5 with 		{\arrow{>}}}, postaction={decorate}] (\offset+\xa-\ra,\yc) -- node[anchor = south] {\color{black} $e^{-i\pi(a+b+c)}$} (\offset+\xstart,\yc);
	\draw[semithick, blue] (\offset+\xb+\ra,\yc) arc [radius=\ra, start angle=0, end angle=-180];
	\draw[semithick, blue] (\offset+\xa+\ra,\yc) arc [radius=\ra, start angle=0, end angle=-180];
	\node at (\offset+\xend,\yc) [anchor=west] {$\times e^{i\pi(b+c)}$};
	
	\draw[decoration={brace,mirror,raise=5pt},decorate]
	(-0.2,\yd) -- node[left=6pt] {$=$} (-0.2,\ye);
	
	\node at (\xstart, 0.5*\yd+0.5*\ye) {$+$};
	
	\draw (\offset+\xstart,\yd) -- (\offset+\xend,\yd);
	\foreach \Point in {(\offset+\xa,\yd), (\offset+\xb,\yd), (\offset+\xc,\yd)}{
		\draw[fill=black] \Point circle [radius=1.2pt];
	}
	\draw[semithick,blue,decoration={markings, mark=at position 0.5 with 		{\arrow{>}}}, postaction={decorate}] (\offset+\xa,\yd) -- (\offset+\xstart,\yd);
	\node at (\offset+\xend,\yd) [anchor=west] {$\times(e^{-i\pi a}-e^{i\pi a})$};
	
	\draw (\offset+\xstart,\ye) -- (\offset+\xend,\ye);
	\foreach \Point in {(\offset+\xa,\ye), (\offset+\xb,\ye), (\offset+\xc,\ye)}{
		\draw[fill=black] \Point circle [radius=1.2pt];
	}
	\draw[semithick,blue,decoration={markings, mark=at position 0.5 with 		{\arrow{>}}}, postaction={decorate}] (\offset+\xc,\ye) -- (\offset+\xb,\ye);
	\node at (\offset+\xend,\ye) [anchor=west] {$\times(e^{i\pi c}-e^{-i\pi c})$};
	
	\foreach \Point in {(\xa,\ya),(\xa+\offset,\yb),(\xa+\offset,\yc),(\xa+\offset,\yd),(\xa+\offset,\ye)}{
		\node at \Point [anchor=north] {$0$};
	}
	
	\foreach \Point in {(\xb,\ya),(\xb+\offset,\yb),(\xb+\offset,\yc),(\xb+\offset,\yd),(\xb+\offset,\ye)}{
		\node at \Point [anchor=north] {$z$};
	}
	
	\foreach \Point in {(\xc,\ya),(\xc+\offset,\yb),(\xc+\offset,\yc),(\xc+\offset,\yd),(\xc+\offset,\ye)}{
		\node at \Point [anchor=north] {$1$};
	}
	
	\end{tikzpicture}
	\caption{\label{fig:deforming1} Deform the integral line of $I_1(z)$ to express it in terms of $I_j(1-z)$.}
\end{figure}
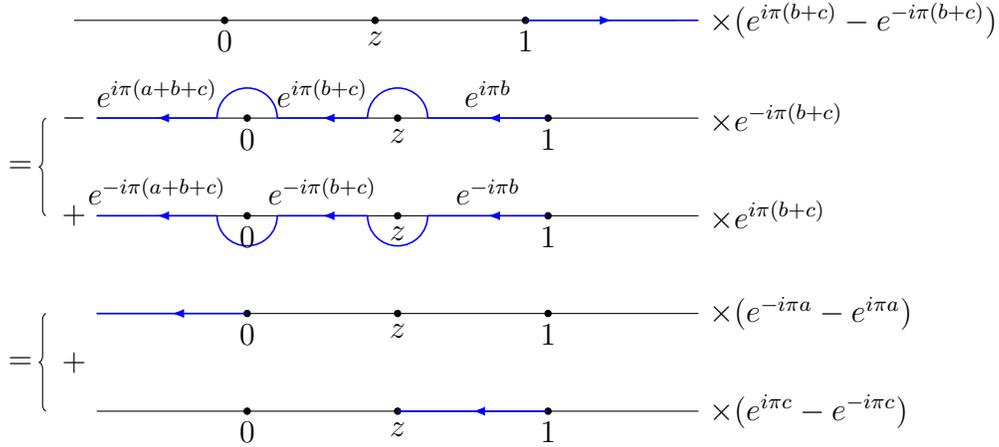

Therefore we only have to know $a_{ij}$. In this problem, we can simply look up the linear transformation formula of Hypergeometric function $F(\alpha,\beta,\gamma,z)$ to get the result. But for more generic choice of operators, conformal blocks are more complicated functions and this method is no longer applicable. The Coulomb gas formalism provides a more generalizable method. We first put the constraints $a>-1, b>-1, c>-1, a+b+c<-1$, where $I_1,I_2$ both have line integral representations. Because $I_1,I_2$ are analytical on the upper and lower have plane, we can try to deform the line. Let us look at $I_1$ here. We can deform the integration contour of $I_1$ in two different ways and multiply them by two different phase factors, as shown by Fig.~\ref{fig:deforming1}. For $I_2$, we can do a similar deformation. The deformed line integrals are exactly $I(1-z)$ with corresponding parameters. And the relation is found to be,
\begin{align}
	I_1(a,b,c;z) = & \frac{s(a)}{s(b+c)} I_1(b,a,c;1-z) - \frac{s(c)}{s(b+c)}I_2(b,a,c;1-z) \\
	I_2(a,b,c;z) = & -\frac{s(a+b+c)}{s(b+c)} I_1(b,a,c;1-z) - \frac{s(b)}{s(b+c)} I_2(b,a,c;1-z)
\end{align}
where $s(a) = \sin \pi a$. Using these two relations, we can work out the monodromy matrix that we write down in the main text. The final result for $G(z,\bar z)$, up to an overall normalization factor $A$, is
\begin{align}
	G(z,\bar z) = A\left[\frac{s(b)s(a+b+c)}{s(a+c)} |I_1(z)|^2 + \frac{s(a)s(c)}{s(a+c)} |I_2(z)|^2 \right],
\end{align}
which is the result that we used in the main text.
Comments on several subtleties are followed:
\begin{itemize}
	\item Here we have to first require $z$ to be real. However, the final result is analytic to $z$ so we can continue $z$ to be a complex number.
	\item Because of the constraints $a>-1, b>-1, c>-1, a+b+c<-1$, during the deformation, we don't have to worry about the integral at infinity or along the small circles around $0,z,1$. However, the final relation is analytical of $a,b,c$ so we can release the constraints and apply the relation to a larger parameter region.
	\item No matter deforming the contour from the upper or lower half plane, we always choose $\arg\omega = \arg (\omega - z) = \arg(\omega-1) = 0$ when doing the integral $\int_1^{+\infty}d\omega$ in $I_1$. For $I_2$, we define it to be $I_2(z) = \int_0^z d\omega \omega^a(1-\omega)^b(z-\omega)^c$ and $\arg\omega = \arg(1-\omega) \arg(z-\omega) = 0$. Only with these choices, can we get the desired phase factors and the final result.
\end{itemize}

\paragraph{Operator Algebra}

If we want to fix the overall normalization factor $A$, we have to use operator algebra, i.e. the coefficient of $WW$ and $VV$ OPE. Because $A$ is not important for the discussion in the main text, we are not going to derive $A$ in details but just go through the logic.

For any two primary fields, the holomorphic part of their OPE can be written as,
\begin{align}
	\phi_{(r_1,s_1)}(z_1) \phi_{(r_2,s_2)}(z_2) = \sum_{r,s} \frac{C_{(r_1,s_1),(r_2,s_2)}^{(r,s)}}{z_{12}^{h_{r_1,s_1}+h_{r_2,s_2}-h_{r,s}}} \phi_{(r,s)}(z_2).
\end{align}
$C_{(r_1,s_1),(r_2,s_2)}^{(r,s)}$ is called the OPE coefficient which is necessary for defining a CFT. And we usually choose the convention that $C_{(r,s)(r,s)}^{(1,1)}=1$. Any correlation function can be calculated by performing the OPE.

Therefore, we have two equivalent ways to calculate the four point function here. One the one hand, we have the fusion rule $\phi_{1,2}\times\phi_{1,2}=\phi_{1,1}+\phi_{1,3}$. When $z_1\rightarrow z_2, z_3\rightarrow z_4$, the four point function can be written terms of the OPE relation as,
\begin{align}
	4pt = \frac{1}{z_{12}^{2h_w}z_{34}^{2h_v}} \left[ 
	1 + C_{(1,2),(1,2)}^{(1,3)}z^{h_{1,3}}\bar z^{h_{1,3}} + ...
	\right],
\end{align}
where we only preserve the leading term expansion of each fusion channel. One the other hand, we can expand $f_{WV}$ obtained from the Coulomb gas formalism at $z,\bar z=0$.
\begin{align}
	f_{WV}(z,\bar z) \approx \# z^{2\alpha_1\alpha_2 + 2h_w} \times c.c. + \# z^{2\alpha_1\alpha_2+2h_w+1+a+c}\times c.c.
\end{align}
with some complicated real coefficient. As a self-consistent theory, these two results should match each other. Simple algebra shows that,
\begin{align}
	2\alpha_1\alpha_2 + 2h_w & = h_{1,3}, \\
	2\alpha_1\alpha_2 + 2h_w + 1+a+c & = 0.
\end{align}
So the exponent indeed matches. Furthermore, by matching coefficients, we can fix the value of $A$.

As a side remark, we can also check the expansion of $f_{WV}$ at $z=\infty$. Here we can do a $1/z$ expansion and get\footnote{
	Here, we use the relation between Hypergeometric function at $z=0$ and $z=\infty$,
	\begin{align}
		F(\alpha, \beta, \gamma, z) = & \frac{\Gamma(\gamma)\Gamma(\beta-\alpha)}{\Gamma(\gamma-\alpha)\Gamma(\beta)} (-z)^{-\alpha} F(\alpha,\alpha-\gamma+1, \alpha-\beta+1, 1/z) \\
		& + \frac{\Gamma(\gamma)\Gamma(\alpha-\beta)}{\Gamma(\gamma-\beta)\Gamma(\alpha)} (-z)^{-\beta} F(\beta, \beta-\gamma+1, \beta-\alpha+1, 1/z),
	\end{align}
	where $-\pi < \arg(-z) < \pi$.
},
\begin{align}
	I_1(z) = \frac{\Gamma(b+1)\Gamma(-1-a-b)}{\Gamma(-a)} (-z)^c + \frac{\Gamma(-a-b-c-1)\Gamma(1+a+b)}{\Gamma(-c)} (-z)^{1+a+b+c} \\
	I_2(z) = \frac{\Gamma(1+a)\Gamma(-1-a-b)}{\Gamma(-b)}(-1)^{-1-a}z^{c} + \frac{\Gamma(1+c)\Gamma(1+a+b)}{\Gamma(2+a+b+c)}(-1)^b z^{1+a+b+c.}
\end{align}
Therefore we have,
\begin{align}
	f_{WV}(z,\bar z) = \# z^{2\alpha_1\alpha_2 + 2h_w + 2\alpha_2\alpha_3 + c} \times c.c + \# z^{2\alpha_1\alpha_2 + 2h_w + 2\alpha_2\alpha_3 +1+a+b+c}\times c.c. 
\end{align}
We can show that the exponents here
\begin{align}
	2\alpha_1\alpha_2 + 2h_w + 2\alpha_2\alpha_3 + c & = -(h_{m,n-1}-h_{1,2}-h_{m,n}), \\
	2\alpha_1\alpha_2 + 2h_w + 2\alpha_2\alpha_3 +1+a+b+c & = 
	-(h_{m,n+1}-h_{1,2}-h_{m,n}),
\end{align}
exactly match the $WV$ OPE, i.e. $\phi_{1,2} \times \phi_{m,n} = \phi_{m,n-1} + \phi_{m,n+1}$.

\section{Integral Representation of Hypergeometric Functions}
\label{app:hypergeo}
The Hypergeometric equation
\begin{align}
	z(1-z) \omega'' + [\gamma - (\alpha+\beta+1)z] \omega' - \alpha\beta \omega = 0
\end{align}
is known to give an analytical solution at $z=0$, the Hypergeometric function 
\begin{align}
	F(\alpha,\beta,\gamma;z) = \sum_{k=0} \frac{(a)_n(b)_n}{(c)_n} z^n.
\end{align}
This result can be easily proved using series expansion method. 

This differential equation can also be solved by the Euler transformation
\begin{align}
	\omega(z) = \int_{\mathcal{C}} dt\,(z-t)^{\mu}\nu(t),
\end{align}
which yields a solution,
\begin{align}
	\omega(z) = A\int_{\calC} dt\,t^{\alpha-\gamma} (1-t)^{\gamma-\beta-1} (z-t)^{-\alpha} + \int_\calC dt \frac{d}{dt}Q, 
\end{align}
where $A$ is an arbitrary constant and the integration over $Q$ is a boundary term,
\begin{align}
	 Q = -A\alpha t^{\alpha-\gamma+1}(1-t)^{\gamma-\beta}(z-t)^{-\alpha-1}.
\end{align}
To get rid of the boundary term, we have to choose the integration path $\calC$ appropriately such that $Q$ takes the same value at the starting and end point. 

When $\Re\gamma>\Re\beta>0$, we can choose the contour to be $\mathcal C = [1,+\infty]$ because $Q$ vanishes at both $t=1$ and $t=+\infty$. So we have,
\begin{align}
	\omega(z) = & A\int_1^{+\infty} dt\,t^{\alpha-\gamma} (1-t)^{\gamma-\beta-1} (z-t)^{-\alpha} \\ \label{eqn:hypergeo1}
	\overset{t\rightarrow1/t}{=} & A'\int_0^{1} dt\,t^{\beta-1} (1-t)^{\gamma-\beta-1} (1-zt)^{-\alpha}, \,\Re\gamma>\Re\beta>0.
\end{align}
Eqn\eqref{eqn:hypergeo1} is uniformly convergent in the vicinity of $z=0$.

If $\text{Re}\gamma>\text{Re}\beta>0$ is not satisfied, we have to resort to other choices. One of such choices is the Pochhammer contour, which is depicted in Fig.~\ref{fig:pochhammer}. We can see that $Q$ as a function of $t$ has branch point at $t=0,1,z,\infty$. The Pochhammer contour circulates each branch point twice but in opposite direction so that we get back to the original Riemann surface, which makes the boundary term vanishes. In the main text, we use the Pochhammer contour integral representation.

\end{document}